\documentclass[journal,transmag]{IEEEtran}

\IEEEoverridecommandlockouts

\usepackage{epsfig}
\usepackage{epstopdf}
\usepackage{times}
\usepackage{algorithm}
\usepackage{algorithmic}
\usepackage{amssymb}

\usepackage{amsmath}
\usepackage{subfigure}
\usepackage{multirow}
\usepackage{url}
\usepackage{eso-pic}
\usepackage{lipsum}
\usepackage{soul}
\usepackage{bbm}
\usepackage{cuted}
\usepackage{makecell}
\usepackage{graphicx}

\usepackage[
  pass,
]{geometry}

\usepackage{array} 
\def\ignore#1\endignore{}
\newcolumntype{h}{@{}>{\ignore}l<{\endignore}} 

\newcolumntype{x}[1]{%
>{\centering\hspace{0pt}}p{#1}}%

\usepackage{latexsym}
\usepackage{multicol}
\usepackage{eurosym}
\usepackage{amsthm}
\usepackage{xspace}
\usepackage{pdfpages}
\usepackage{verbatim}
\usepackage{bigints}

\newenvironment{sketch}{{\noindent \it Sketch of Proof:~}}

\newtheorem{proposition}{Proposition}

\newtheorem{lemma}{Lemma}

\newcommand{\powerset}{\raisebox{.15\baselineskip}{\Large\ensuremath{\wp}}}

\makeatletter
\def\@copyrightspace{\relax}
\makeatother

\begin{document}
\graphicspath{{./Figures/}}
\title{ONETS: Online Network Slice Broker\\From Theory to Practice}

\author{Vincenzo~Sciancalepore,~\IEEEmembership{Senior Member,~IEEE,}
        Lanfranco~Zanzi,~\IEEEmembership{Associate Member,~IEEE,}\\
        Xavier~Costa-Perez,~\IEEEmembership{Senior Member,~IEEE,}
        and~Antonio~Capone,~\IEEEmembership{Fellow,~IEEE}
\IEEEcompsocitemizethanks{\IEEEcompsocthanksitem V.~Sciancalepore and L.~Zanzi are with NEC Laboratories Europe GmbH, Kurf\"ursten-Anlage 36, 69115 Heidelberg, Germany. E-mail: \{name.surname\}@neclab.eu.  
\IEEEcompsocthanksitem X.~Costa-P\'erez is with NEC Laboratories Europe GmbH., Kurf\"ursten-Anlage 36, 69115 Heidelberg, Germany, and i2CAT Foundation and ICREA, Carrer del Gran Capità 2, 08034 Barcelona, Spain, Email: xavier.costa@neclab.eu.
\IEEEcompsocthanksitem A.~Capone is with Politecnico di Milano, Dipartimento di Elettronica, Informazione e Bioingegneria, Piazza L. da Vinci 32, 20133 Milan, Italy. E-mail: antonio.capone@polimi.it.}
\thanks{Manuscript received XXX XX, XXXX; revised XXX XX, XXXX.}}

\maketitle

\begin{abstract}
Network slicing allows mobile network operators to open their physical network infrastructure platform to the concurrent deployment of multiple logical self-contained networks, i.e., \emph{network slices}. In this paper we propose and analyze \emph{ONETS}: an Online NETwork Slicing solution that ($i$) builds on the budgeted lock-up multi-armed bandit mathematical model and properties, ($ii$) derives its analytical bounds in our proposed extension for network slicing, ($iii$) seamlessly integrates into the 3GPP architecture, ($iv$) proves its feasibility through a proof-of-concept implementation on commercial hardware considering three network slices and ($v$) allows for the design of a low-complexity online network slice brokering solution that maximizes multiplexing gains. 


\end{abstract}


\psfull

\setlength{\textfloatsep}{18pt}

\thispagestyle{empty}	

\section{Introduction}
\label{s:intro}

Mobile networks are a key element of today's society, enabling communication, access and information sharing. However, as cellular networks move from being voice-centric to data-centric, operators revenues are not able to keep pace with the predicted increase in traffic volume. Such pressure on operators return on investment has pushed research efforts towards designing novel mobile network solutions able to open the door for new revenue sources.

The emerging \emph{network slicing} paradigm provides new business opportunities by enabling mobile operators to open their network infrastructure to multiple \emph{tenants}, i.e., slice owners, with very diverse requirements. The availability of this vertical market multiplies the monetization opportunities of the network infrastructure as ($i$) new players may come into play (e.g., automotive industry, e-health, etc.) bringing in new business opportunities, and ($ii$) a higher infrastructure capacity utilization can be achieved by admitting as many network slice requests as possible while exploiting stochastic multiplexing gains obtained from the shared multi-tenant infrastructure. With network slicing, different services can be provided by different network slice instances. Each of these instances consists of a set of virtual network functions that run on the same infrastructure with a tailored orchestration.

However, network slicing introduces new challenges that need to be addressed in order to be adopted in practice. A trade-off has to be considered between a \emph{fully shared} mobile network among tenants (with shared functions and resources) and an \emph{isolated slices} one (with dedicated functions and resources only). In this context, a network slice broker solution is desirable, acting  as an arbitration entity in charge of satisfying heterogeneous slice requirements from tenants while at the same time guaranteeing the most efficient use of the infrastructure resources~\cite{Slicing2018CoNEXT}. The network slice broker concept has been previously considered in~\cite{samdanis2016network}. In this paper we build on this concept and design an \emph{online} network slice brokering solution that complies with the novel 3GPP Network Slicing architecture development (described in Section~\ref{s:framework}). 

The objective of our \emph{ONETS} solution is to design an efficient online network slice broker that by analyzing past network slicing information maximizes future revenues deriving from the satisfaction of network slice requests, while exploiting multiplexing gains coming from the overbooking of resources and increasing the overall system utilization. In particular, we provide 
($i$) a novel decisional model addressing the ``exploration vs exploitation'' dilemma, dubbed as Budgeted Lock-up Multi Armed Problem (BLMAB),
($ii$) a detailed analysis of such a class of problems, including specific exploitable features to design a feasible efficient solution maximizing multiplexing gains,
($iii$) multiple variants of the proposed solution accounting for complexity and optimality properties along with performance upper bounds, 
($iv$) an exhaustive simulation campaign with synthetic traces to get an indication of the expected benefits in large-scale scenarios and
($v$) a proof-of-concept implementation using commercial hardware to prove the feasibility of our solution considering three network slices: enhanced Mobile BroadBand (eMBB) for Guaranteed Bit Rate (GBR), eMBB for Best Effort (BE), and Public Safety.


The remaining of the paper is organized as follows. Section~\ref{s:framework} shows the architectural requirements of our solution. Section~\ref{s:related} lists the state-of-the-art solutions. Section~\ref{s:design} presents the class of decisional problems by deriving tight bounds and insightful properties. Section~\ref{s:online} proposes different online algorithms trading-off between complexity and sub-optimality. Section~\ref{s:validation} shows experimental results to prove our findings with synthetic network scenarios. Section~\ref{s:implement} presents our proof-of-concept built on top of commercial equipment and corresponding results. Finally, Section~\ref{s:concl} provides our conclusions.

\section{Network Slicing in 3GPP} 
\label{s:framework}

\begin{figure} [!t]
\centering
	\includegraphics[clip, trim = 2cm 10cm 1.5cm 6.5cm, width=0.5\textwidth]{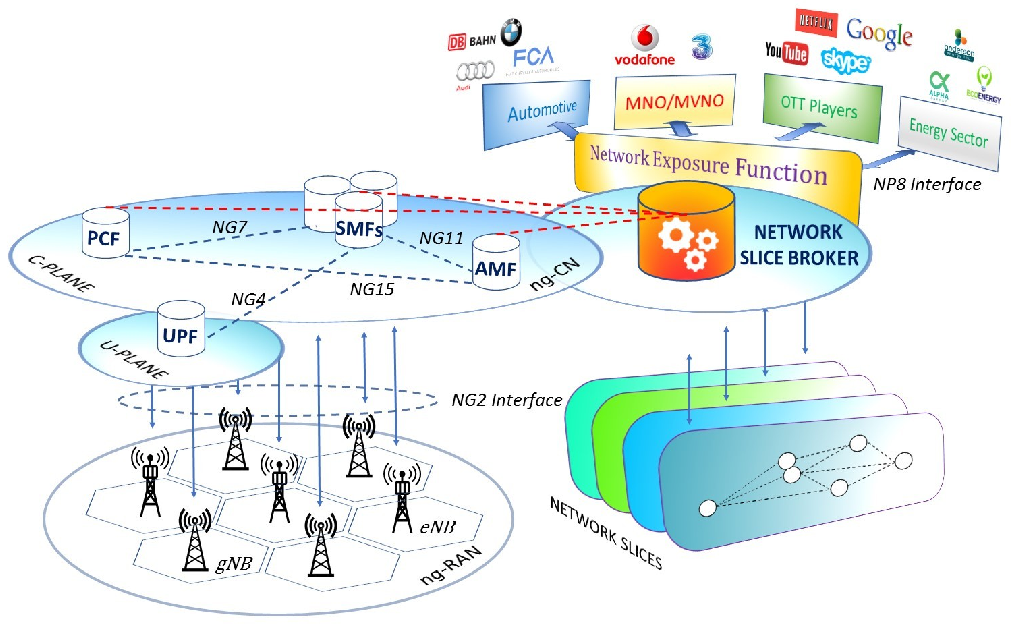}
	\caption{3GPP Network Slicing Architecture}
	\label{fig:standard_infrastr}
\end{figure}

3GPP has defined a novel network architecture for network slicing support. In particular, the 3GPP working group SA2~\cite{tr23501} has already defined the basis for building an evolved core network infrastructure managing multiple slices on the same network infrastructure. The envisioned architecture is depicted in Fig.~\ref{fig:standard_infrastr} which clearly differentiates between control plane (C-Plane) and user plane (U-Plane). In the control plane, new components are introduced to ($i$) manage user authentication and registration (AMF), ($ii$) support multiple connection sessions (SMF) and ($iii$), instruct different routing policies (PCF). On the other hand, the user plane is unified into a generic function (UPF) managing distinct data networks (DNs) through the next-generation-Radio Access Network (ngRAN). This new architecture allows for an easier network functions virtualization and thus, flexible multi-tenant deployments. RAN nodes (and functions) are virtualized and flexibly chained to provision end-to-end RAN slices with a dedicated SMF. Interestingly, AMF (and PCF) can be shared among multiple slices when presenting service requirements commonalities. 

Based on this architecture, the Network Exposure Function (NEF) can be used as a direct interface between the mobile network operator and the network slice tenants to access the virtualized network functions. NEF is envisioned to expose a list of available slice templates defining specific functions to be instantiated for given service requirements. Network slice request coming through the NP8 interface will then indicate the requested slice template based on the available ones. At this point, an arbitration entity is needed to grant (or deny) network slice requests. 
Once a network slice request is granted, a Network Slice Selection Assistance Information (NSSAI) indicator is propagated through all network components and advertised to incoming UEs through the RAN. Based on the NSSAI, the AMF will select the SMF and a network slice will be successfully installed. Associated UEs might then indicate in the RRC signaling the NSSAI to be used for serving its traffic.

In Fig.~\ref{fig:standard_infrastr} we depict the proposed location of the arbitration entity in charge of granting or denying network slice requests, referred in the paper as \emph{Network Slice Broker}. In the following we review the state-of-the-art of network slicing solutions available in the literature to meet the functionality required by an online  network slice broker.

\section{Related Work}
\label{s:related}
Network slicing is currently a very hot topic in the 5G research community given its business relevance~\cite{alliance2015description} and the recent definition by 3GPP of the architecture \cite{23.799}. Network virtualization though is not a new topic and several solutions supporting virtualization of network resources have been already designed to optimize network resource utilization and improve QoE.

%
%
%
The authors of~\cite{foukas_orion} focus on the RAN domain, and propose an efficient sharing of the radio resources which performs adaptive provisioning and isolation of network slices by means of a bi-level scheduling approach.
From a radio physical channel point of view, the work of~\cite{Oro_Infocom_2019} suggests accurate inter-cell scheduling strategies to mitigate the interference effects coming from multiple base stations serving the same sets of slices, thus improving the overall channel conditions and slice performances. As a drawback, this scheme would introduce significant synchronization overhead, whose impact on network management strategies should be investigated.
The goal of a network slice broker is to coordinate slice requests as well as to ensure the dynamic provisioning of networking resources in an efficient way. In~\cite{RL-NSB}, the slicing admission control and scheduling tasks are introduced by means of an NP-hard optimization problem formulation. The authors suggest a reinforcement learning-based approach which exploits resource utilization and traffic predictions to improve the overall admission control decision process. Their analysis details on the achievable benefits regarding the number of admitted network slice requests and the system capacity utilization.
Similarly, in~\cite{LACO}, the authors present a reinforcement learning-based RAN orchestration solution focusing on latency constraints. Their framework allows for adaptive resource slicing decisions without prior knowledge of slice-specific traffic demands or channel statistics.
\cite{jiang2016network} suggests a heuristic-based prioritized admission control mechanism. The novelty here stands in the possibility to adapt the algorithm to both inter- and intra-cell admission control problem. The admission procedure of a new UE belonging to the same slice is done taking in account the current traffic load and the available resources. Only if there are enough resources to guarantee and satisfy at least the requirements on a predefined minimum data rate, the procedure will be successfully ended.
\cite{Sciancalepore2017Mobile} mathematically analyses the admission control issue for network slicing and proposes an algorithmic solution by applying machine learning concepts. However, it relies on offline approach, which gathers several network slice requests within a fixed time window while selecting (some of) them as so to maximizing the overall network utilization.
%
%
%
As pointed out in~\cite{jiangnetwork} slicing implies that each 5G slice needs to have its own set of allocated resources and this aspect introduces a novelty in the management of network resources in mobile systems. Indeed, in previous generations of mobile networks, the resources to be assigned were mainly radio resources, while in 5G networks it is commonly accepted that this will impact also on the core network resources splitting. This easily matches with the research effort in Network Function Virtualization (NFV) and Software Defined Network (SDN) fields toward the deployment of a modular and flexible 5G network, as discussed in~\cite{richart2016resource,5GCrosshaul}. 
\cite{choyi2016network} presents a framework for enabling negotiation, selection and assignment of network slices for requesting applications in future 5G networks. Based on different QoS Class Indicators (QCIs), different virtualized networks or network slices are selected and assigned to users demanding for a specific service. Subsequently, static or dynamic routing mechanisms are used to treat data packets according to the QCI and security requirements and to flexibly select network functions and service function paths through a NS. In the same context, \cite{Sciancalepore2017Optimising} introduces a practical admission control solution for network slicing pursuing at maximizing the overall network revenues following a simple pricing model for RAN slices.
%
A slice selection function is a key element in the future core network architecture. \cite{sama2016service} attempts to fill the gap between the emerging new service requirements in terms of performance and efficiency and the possible realization of the concept by proposing a new slice selection mechanism allowing the UE to connect to multiple slices based on service types.

To the best of our knowledge, the work presented in this paper is the first of its kind that aims at designing and evaluating an automated online network slice broker while pursuing the network slicing multiplexing gain maximization.

\section{System Design}
\label{s:design}

We consider a telecom service provider (hereafter dubbed \emph{Operator}) making available its own infrastructure to external network tenants, for e.g. vertical industries. The operator owns the entire access infrastructure with a fixed transmission capacity $C$\,\footnote{In our work, we consider the network slicing paradigm uniformly applied across the network, as currently stated by 3GPP~\cite{23.799}.}. 
The available tenants set $\mathcal{I}$ is known a-priori, as each external domain must subscribe beforehand to submit network slice requests and get access to the infrastructure.
Each tenant $i \in \mathcal{I}$ can request a network slice $s \in \mathcal{S}$ best matching the network characteristics required amongst available network slice templates (NSTs). Such templates are considered as fixed based on the operator's business model~\cite{alliance2015description}. 
Each network slice request $s=\{R^{(s)}; L^{(s)}\}$ comprises an amount of physical wireless resources, $R^{(s)}$, to be assigned to tenant users within the network slice $s$, and a time duration $L^{(s)}$ expressed in seconds. Such parameters might be tailored to particular services and might be modified according to new service requirements.
%
Indeed, the network slice requirements define the Service Level Agreement (SLA) between the operator and the infrastructure tenant. Any failure to comply with such SLA might result in undesired penalties from the operator side.

In the next subsection, we first introduce the objective of the operator to increase its revenue opportunities while keeping SLA violations negligible, and then we detail the system model behind such problem as well as our proposed solution.

\subsection{System Model}
Let us consider a set of $T$ time instants denoted by $\mathcal{T}=\{1 ,\dots,T\}$. Each tenant $i \in \mathcal{I}$ asks for a network slice template $s$ at time instant $t \in \mathcal{T}$, modeled with the variable $r_i^{(s)}(t)$. The requests distribution is obtained from an i.i.d. random variable, namely inter-arrival time $\Delta t$, exponentially distributed with rate $\phi_i$. Inter-arrival rates, $\phi_i$, are drawn from a Pareto distribution with mean $\rho$ and standard deviation $\zeta$, determining the level of heterogeneity\,\footnote{Note that this assumption relies on the traffic flow behavior, as suggested in~\cite{trafficAnalysis}. However, as shown in Section~\ref{s:ubEUCB}, it might be relaxed to bring interesting findings.}.

The tenant selects a network slice template $s=\{R^{(s)};L^{(s)}\}$ to be issued with the network slice request. In our analysis, this choice is taken based on 
tenants network slice requirements to efficiently drive the slice selection process. We also assume that each tenant can only ask for a single network slice at a given time and, tenants can be granted only a single network slice request, i.e., multiple network slices assigned to the same tenant cannot overlap in time, $t \geq t_{-1} + L^{(s)}$, if the slice template $s$ has been granted in the previous request $r_i^{(s)}(t_{-1})$\,\footnote{This assumption makes tractable the analysis in Section~\ref{s:blmab}. However, it can be easily extended by assuming tenants asking for multiple slices at the same time as distinct virtual tenants.}.  

Without loss of generality, we express network slice request as $r_{i,t}^{(s)} = \{ R^{(s)}_{i,t};L^{(s)}_{i,t}\}$ and the problem as follows: Upon receiving a network slice request, \emph{the operator decides whether to accept or reject it in an online fashion, pursuing the objective of network slicing multiplexing maximization while still honouring the agreed  guarantees (SLAs) for previously granted network slice requests}.

\subsection{Online Decisions:\\ Exploration vs Exploitation dilemma}
Once a network slice request is received, the operator might decide (at runtime) to accept or reject it, based on different factors: ($i$) deterministic aspects and ($ii$) stochastic components. The former group includes the set of requirements for network slices  currently running and the total available system capacity. The latter comprises random tenant choices when issuing the slice request, upcoming network slice requests and real network utilization within an allocated network slice.
The operator can decide to allocate an incoming network slice request fitting into available network resources. However, this might prevent future network slice requests from being accepted, though they can further boosting the network utilization. This may negatively affect the overall online process, as the current selection decision is strongly tied to future admissions. While deterministic considerations can be efficiently taken into account, stochastic features need advanced mechanisms to drive the system towards a near-optimal system behavior. 

\subsubsection*{Multi-Armed Bandit Model}
When dealing with an online decision process, a plethora of mathematical tools and practical schemes helps to bound the space of solutions and provides affordable and sub-optimal results~\cite{Vermorel2005,proutiereINFOCOM2014,Kleinberg2008}. We focus in this dissertation on a subset of online algorithms considering sequential decisions with limited information. 
In particular, we envisage a gambler facing diverse game options to play, resulting in different gains. The player must sequentially select the best option (i.e., the tenant slice request) in order to maximize the profit (i.e., overall system utilization). This results in the fundamental \emph{exploration vs exploitation} lemma during sequential experiments: The gambler needs to balance the exploitation of known tenant slice requests that paid well in the past and the exploration of upcoming tenant slice requests that might pay even more.

Sequential allocation problems fully match Multi-Armed Bandit models (MABs)~\cite{survery_mab}. In particular, the fundamental problem formulation is obtained from a casino use-case, where a gambler faces with multiple slot machines (i.e., bandits). Slot machine return unpredictable revenues obtained through unknown statistical functions. The gambler can play one coin at once to ($i$) observe the profit behavior of unexplored slot machines or ($ii$) keep playing with the one providing (in the previous rounds) the best profit. The final objective is to maximize the overall profit after playing a finite number of rounds. While this model has been fully investigated, our problem needs substantial improvements to be treated as a novel variation of MAB.

\subsection{A Budgeted Lock-up Multi-Armed Bandit Problem (BLMAB)}
\label{s:blmab}
We build on top of the basic MAB our problem formulation by introducing three fundamental MAB variations: ($i$) multi-plays, ($ii$) limited budget and ($iii$) lock-up periods. 

Let us consider each tenant $i$ as a bandit that, if pulled at round $t$, returns a certain reward $\eta_{i,t}$. Multiple tenants can ask (simultaneously) a network slice request, hence the gambler may play  multiple bandits at the same round, i.e., she may select multiple tenants to be granted at the same time. To avoid a trivial solution pulling down all arms to maximize the total revenue, our formulation introduces a cost function for selecting those bandits: At every round, the player needs to select a batch ($K \geq 1$) of arms whose cost lies within the available budget. 

We define a cost function $\lambda_{i,t}$ as the number of resource blocks (PRBs) used within the round $t$. The total budget available for selecting slice requests is $C$, that is the total available system capacity. This defines our admissibility region, as more than available PRBs cannot be allocated.

Since different network slice requests might come at different times and occupy the resources for fixed time intervals, we need to modify our model accordingly. Basically, we account for the case when a pulled-down arm does not return any payoff, as it directly translates into the case of a tenant selected to be granted at time $t$ but not interested in issuing network slice requests at that time. Although this behavior looks counterintuitive, it provides the effective means for deeply learning the tenant behaviors (by considering the frequency of network slice requests and the real slice utilization) and predict future requests.

On the other side, a network slice request $r^{(s)}_{i,t}$ granted for tenant $i$ at time $t$ must be considered active for the next rounds $t$ until $L^{(s)}$ expires. Both features are taken into consideration by introducing the concept of lock-up periods: On each round $t$, if a lock-up period is running, the gambler must select the same arm as in the previous round (however, the gambler can still select multiple arms in that round).

\begin{figure} [!t]
\centering\includegraphics[clip,trim = 3.5cm 11.5cm 3.5cm 1cm, width=0.48\textwidth]{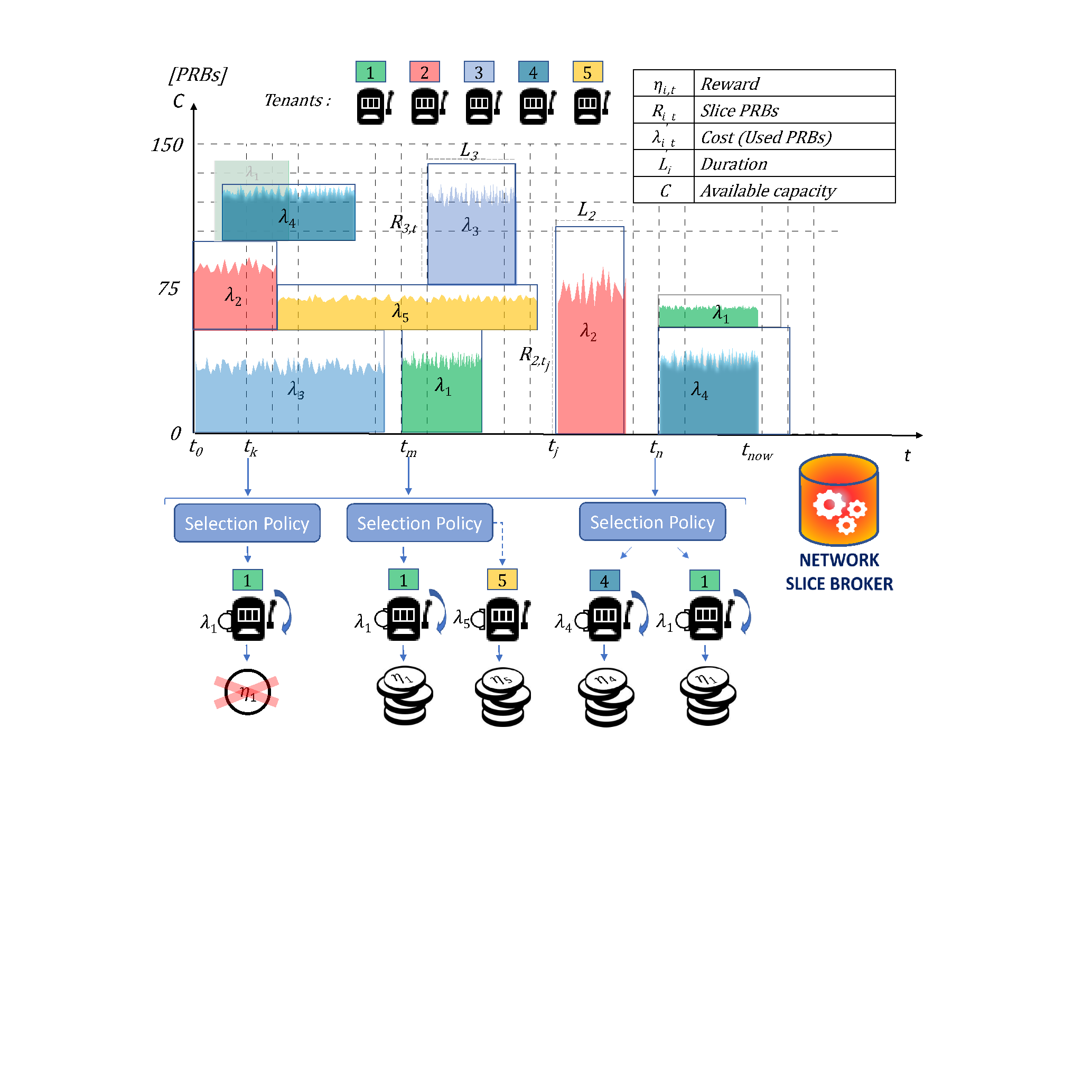}
	\caption{BLMAB problem example applied to an online network slicing brokering process.}
	\label{fig:ModelResume}
\end{figure}

A self-explained example is depicted in Fig.~\ref{fig:ModelResume}, where three different rounds are highlighted ($t_k$, $t_m$ and $t_n$). In the first case, the selection policy decides to grant tenant $1$ but no network slices are issued from this tenant, returning no rewards. In the second case, the selection policy grants tenant $1$ asking for a network slice (and hence returning a reward). It also reselects the previous tenant ($5$) as its lock-up is still running. In $t_n$ the selection policy can select tenant $4$ and $1$ getting rewards, as they ask for new network slices.
The case of a traffic request coming from a tenant and not pulled by the broker, will simply force the tenant to perform a second request in a subsequent time instant.

All in all, we can express the \emph{reward} $\eta_{i,t}$ as the following
\begin{equation}
\label{eq:reward}
\eta_{i,t} = \alpha \frac{R^{(s)}_{i,t}}{C} + (1-\alpha)\frac{R^{(s)}_{i,t}-\lambda_{i,t}}{R^{(s)}_{i,t}},
\end{equation}
where $\alpha \in [0,1]$ is a weight parameter, and it holds 
\begin{equation}
\label{eq:assumpt}
\lambda_{i,t} \leq R^{(s)}_{i,t} \leq C, 
\end{equation}
so that no negative values are obtained, i.e., $\eta_{i,t} \in [0,1]$.
Specifically, the reward accounts for the global amount of resources asked within the slice request (left side of Eq.~\eqref{eq:reward}) as well as for the multiplexing gain, i.e., the ratio between what has been really used and what is being asked (right side). The rationale behind relies on the concept of discovering bargains. Tenants underutilizing assigned resources are preferred with respect to the ones fully using them\,\footnote{In this paper we assume uniform pricing for slice resources. 
}. 

Operators may reuse spare resources to allocate additional network slice requests so as to increase the network utilization (and, in turn, to increase overall system revenues). Additionally, we also take into consideration the total amount of PRBs, as operator might prefer to assign resources to tenants asking (and paying) for more resources, again pursuing the system utilization maximization. $\alpha$ provides a trade-off between those different metrics. 
However, in case of monitoring information not available, the second term will be null and the reward is equal to $\eta_{i,t} = \alpha \frac{R^{(s)}_{i,t}}{C}$. Please note that, when tenant(s) is(are) selected to be granted with no pending slice requests, the total amount of PRBs asked is $R^{(s)}_{i,t} = 0$ resulting in reward $\eta_{i,t} = 0$. Notably, as explained before, the reward expressed in Eq.~\eqref{eq:reward} indirectly accounts also for the tenants behavior, such as the inter-arrival time between consecutive slice requests.
Every round $t$, the operator selects tenants to be granted for being allocated through a set of binary actions $\mathcal{A}_t$. We can now formulate our problem as the following.

\vspace{2mm}\noindent \textbf{Problem}~\texttt{ONLINE-SLICING}:
\begin{equation*}
\label{pr:nw-slicer}
\begin{array}{lll}
\text{maximize}  & \sum\limits_{t\in{\mathcal{T}}} \sum\limits_{i\in{\mathcal{A}_t}} \eta_{i,t}&\\
\text{subject to } & \sum\limits_{i\in{\mathcal{A}_t}} \lambda_{i,t} \leq C, \quad\forall t\in\mathcal{T}; & \text{\footnotesize{:budget}}\\
				   & a_{i,t} \in \mathcal{A}_t \subseteq \{0,1\}^{|\mathcal{I}|}, \quad\forall t\in\mathcal{T}; &\text{\footnotesize{:multi-plays}}\\
                   & a_{i,t} \geq a_{i,t_{-1}}\mathbbm{1}(t-t_i^{\text{START}} \leq L^{(s)}_i); & \text{\footnotesize{:lock-up}}
\end{array}
\end{equation*}
where the last constraint (:lock-up) imposes to select the same arm as in the previous turn, if the lock-up period is still running. With some abuse of notation, we denote $t_i^{\text{START}}$ as the round when slice request has been allocated for tenant $i$ and $\mathbbm{1}(\cdot)$ is an indicator function providing value $1$, if the condition in brackets is satisfied.

\begin{lemma}
\label{lemma:blmab}
A network slice online brokering can be mapped onto a multi-armed bandit (MAB) model with novel variations, such as ($i$) multi-plays, ($ii$) limited budget and ($iii$) lock-up periods. Therefore Problem~\texttt{ONLINE-SLICING} falls into a new class of MAB problems, namely Budgeted Lock-up Multi-armed Bandit problems (BLMAB).
\end{lemma}
\begin{sketch}
Problem~\texttt{ONLINE-SLICING} is a specific instance of MAB, as stated in Lemma~\ref{lemma:blmab}. We apply a reduction to Problem~\texttt{ONLINE-SLICING} by assuming that only one tenant $i$ can be selected every round $t$, i.e., $\sum_{i\in\mathcal{I}} a_{i,t} \leq 1, \forall t\in\mathcal{T}$. Additionally, we assume that each network slice request lasts a single round, i.e., $L^{(s)}=t-t_{-1}$. This implies that the first constraint (:budget) is always satisfied due to Eq.~\eqref{eq:assumpt}. Therefore, we can state that Problem~\texttt{ONLINE-SLICING} can be easily reduced with polynomial reductions to a MAB problem.
\end{sketch}
\qed

Lemma~\ref{lemma:blmab} states that every advanced algorithm solving BLMAB also provides solutions to our network slicing online brokering problem, as we will show in Section~\ref{s:online}.

\begin{proposition}
\label{Prop:hardness}

The BLMAB problem is more complex than the stochastic multi-armed bandit problem, where the forecaster can switch arms at every round. Therefore, a regret lower bound of the stochastic MAB is also a regret lower bound of the BLMAB problem, considering the same number of rounds.

\end{proposition}

Proposition~\ref{Prop:hardness} suggests that we can use the lower bound suggested for MAB to provide a reference point to our mechanism. As we show in the next section, we build on top of such a lower bound to further provide findings on tight bounds for this given class of decision policies.  

\subsection{Regret Lower Bound}
\label{s:regret_lb}

The performance of MAB algorithms can be measured by given metrics, namely regret. The regret denotes the difference in terms of rewards between actions played according to an arbitrary selection policy and the optimal selection policy aware of all reward distributions~\cite{lai1985}. 
Let us consider a player selecting a set of arms $\mathcal{K} \subseteq \mathcal{I}$ every round, such that the budget constraint is fulfilled (refer to constraint :budget of Problem~\texttt{ONLINE-SLICING}). Each arm $i \in \mathcal{K}$ is associated with a univariate known probability density function $f(x,\theta_i)$ , where $x$ is the random variable, all $\theta_i\in\Theta$ are unknown parameters and $\Theta$ represents the set of the reward distributions. Every time arm $i$ is pulled, it returns a reward $\upsilon_i$ drawn from $f(x,\theta_i)$ such that $\mu(\theta_i)$ is the mean of $\upsilon_i$. Let us now consider $\pi = \{\pi(t)\}_{t=1}^T$ as an arbitrary selection policy. The optimal cumulative reward is provided by selecting $|\mathcal{K}|$ arms with the highest reward, i.e., $\{i|i\in \{\sigma_1,\sigma_2,\cdots,\sigma_{|\mathcal{K}|} \}\}$, where $\sigma$ is a permutation vector of $\mathcal{I}$ following a reward decreasing order. Mathematically, it holds that the regret is obtained as 
\begin{equation}
\label{eq:regret}
\begin{array}{ll}
R_T^{\pi}(\Theta) & = T\sum\limits_{i=1}^{|\mathcal{K}|}\mu(\theta_{\sigma(i)})-\mathbb{E}_{\pi}[\sum\limits_{t=1}^T\upsilon_{\pi(t)}(t)]\\
				  & = T\sum\limits_{i=1}^{|\mathcal{K}|}\mu(\theta_{\sigma(i)})-\sum\limits_{i=1}^{\mathcal{I}} \mu(\theta_i) \mathbb{E}[W_i(T)];
\end{array}
\end{equation}
where $W_i(t)$ is the number of rounds arm $i$ has been pulled down till time instant $t$, and $T$ represents the last instant of the time period $\mathcal{T}$.
If the selection policy is \emph{uniformly good}, then $R_T^{\pi}(\Theta) = o(T^{a}), \forall a>0$, and, in turn, it holds that
\begin{equation}
\lim_{\text{T}\to\infty} \sum\limits_{i=1}^{\mathcal{I}} T^{-1}\mu(\theta_i) \mathbb{E}[W_i(T)] = \sum\limits_{i=1}^{|\mathcal{K}|}\mu(\theta_{\sigma(i)}). 
\end{equation}

Therefore, we can express the lower bound of the regret for any uniformly good policy as the following
\begin{equation}
\lim_{\text{T}\to\infty}\inf \frac{R_T^{\pi}(\Theta)}{logT} \geq \!\!\!\!\sum\limits_{i:\,\, \mu(\theta_i)<\mu(\theta_{\sigma(|\mathcal{K}|)})} \!\!\!\!\frac{\mu(\theta_{\sigma(|\mathcal{K}|)})-\mu(\theta_i)}{H(\theta_i,\theta_{\sigma(|\mathcal{K}|)})},
\end{equation}
where $H(\theta_u,\theta_v)=\mathbb{E}\,\log(\frac{f(x,\theta_u)}{f(x,\theta_v)})$ is the relative entropy of one statistical distribution with respect to the other, characterized by $\theta_u$ and $\theta_v$, respectively.



\section{Online Network Slice Broker}
\label{s:online}

Although online network slicing brokering solutions can benefit from being fully customized and not requiring human interventions, a proper design needs advanced algorithms to achieve near-optimal performance. We focus on different classes of solutions, which are explained and analyzed next. 

Algorithms proposed for classical multi-armed bandit (MAB) problems trade off the exploitation of good payoffs with exploration of unknown rewards. This makes such solutions robust and practical resulting in $O(log T)$ as expected cumulative regret. When important variations are considered, the overall complexity might be perturbed by additional factors, e.g., the lock-up time periods.

\subsection{Index-based policy algorithms}
The first class of selection algorithms computes an index per arm. Such an index is updated based on the set of arms already selected in the past, including information regarding the total time elapsed. By doing that, it guarantees the uniformly goodness property (as explained in Section~\ref{s:regret_lb}), which might optimally bound the performance.
We consider the classical Upper Confidence Bound (UCB) solution, enhanced to address our BLMAB problem. 

\begin{algorithm}[t!]
\caption{\small{\textbf{eUCB:} Selection algorithm $\pi$ to select the next batch of arms to pull down while guaranteeing a fixed budget.}}
\label{alg:ucb}
\algsetup{indent=1em}
\begin{footnotesize}
\begin{algorithmic}
	\STATE {\bf Input: }$ \mathcal{I}, T, C$
	\STATE {\bf Initialization: } $W_i = 0, \bar\theta_i(0) =0 \in \bar\Theta, \forall i\in\mathcal{I}, \mathcal{L} \leftarrow \emptyset$
	\STATE {\bf Procedure}
	\end{algorithmic}
	\begin{algorithmic}[1]
\FORALL {$i \in \mathcal{I}$}
	\STATE GET $\upsilon_i$	
    \STATE UPDATE $\bar\theta_i(0)$	
    \STATE $W_i = W_i +1$
\ENDFOR
\FORALL {$t \in T$}
	\STATE $\hat\theta_i(t) = \bar\theta_i(t) + \sqrt{\frac{2 \log t}{W_i}}$
    \STATE $\mathcal{R}\!\! =\!\! \{i\}\!\! \leftarrow$ Problem~\texttt{D-ONLINE-SLICING}$(C,\mathcal{L}(t),\hat\Theta(t))$
	\FORALL {$i \in \mathcal{R}$}
	    \STATE GET $\upsilon_i$	
	    \STATE UPDATE $\bar\theta_i(t)$	
	    \STATE $W_i = W_i +1$
	\ENDFOR
    \STATE UPDATE $\mathcal{L}(t)$
\ENDFOR
\end{algorithmic}
\end{footnotesize}
\end{algorithm}

\subsubsection{Enhanced-UCB (eUCB)}
Several works address the UCB solution considering different variations. We focus on the classical one proposed in~\cite{lai1985} where we collect at every ``attempt'' the reward obtained from each arm and infer the mean of the statistical distribution $\bar\theta_i$. Without loss of generality, we assume that the density function is parameterized with its mean, i.e., $\mu(\theta_i) = \theta_i$. Clearly, the larger the number of attempts, the more accuracy on the distribution information. 
To avoid the negative influence of random effects, the authors include an additional term $G_{\text{ALG}}$ to give more weight to empirical distribution means obtained in longer time windows, than averages obtained in shorter time windows. Also, to address starvation issues, each empirical distribution is weighted with the number of times that an arm has been selected. Index per arm $i$ is formulated as $\hat\theta_i(t) = \bar\theta_i(t) + G_{\text{ALG}}$, where $G_{\text{ALG}}=\sqrt{\frac{2 \log t}{W_i(t)}}$, and $\bar\theta_i(t)$ is the empirical distribution mean until time $t$.

This could be envisioned as a training session where tenants subscribing for network slicing operations, may express their interests on given slice template and, in turn, result in different initial fictitious rewards $\upsilon_i \sim f(x,\theta_i)$. 


It can be noticed that line $8$ optimally solves an instantaneous version of the problem ~\texttt{ONLINE-SLICING}, assuming as input only one time instant $t$, the total budget, the lock-up time windows currently running and the empirical mean of reward distributions retrieved until that time. Please note that the instantaneous version significantly reduces the complexity of our problem, pursuing only the punctual reward maximization at time $t$ rather than the cumulative reward over time window $T$. Output of this problem is a set of arms indexes that are promptly selected in lines $9-13$. Line $14$ updates the status of current lock-up periods for next selections.

\begin{algorithm}[t!]
\caption{\small{\textbf{ONETS:} Selection algorithm $\pi$ to select $K$ arms to pull down while guaranteeing a fixed budget.}}
\label{alg:ucb_k}
\algsetup{indent=1em}
\begin{footnotesize}
\begin{algorithmic}
	\STATE {\bf Input: }$ K, \mathcal{I}, T, C$
	\STATE {\bf Initialization: } $B\! = 0; n\! = 0 ; W_i\!=0, \bar\theta_i(0)\! =0\!\in\!\bar\Theta , \forall i\in\mathcal{I}; \mathcal{L} \leftarrow \emptyset$
	\STATE {\bf Procedure}
	\end{algorithmic}
	\begin{algorithmic}[1]
\FORALL {$i \in \mathcal{I}$}
	\STATE GET $\upsilon_i$	
    \STATE UPDATE $\bar\theta_i(0)$	
    \STATE $W_i = W_i +1$
\ENDFOR
\FORALL {$t \in T$}
	\STATE $\hat\theta_i(t) = \bar\theta_i(t) + \sqrt{\frac{2 \log t}{W_i}}$
    \WHILE{$n\leq K \land B \le C$ }
    	\IF{$\mathcal{L}(t) \neq \emptyset$}
        	\STATE $\hat{i} \leftarrow \mathcal{L}(t)$
        \ELSE
    		\STATE $\hat{i}: \arg\max\limits_{\mathcal{I}\setminus\mathcal{L}} \hat\theta_i(t)$
        \ENDIF
        \IF{$B + \lambda_{\hat{i}} \leq C$}
			\STATE $\mathcal{R} \leftarrow \mathcal{R} \cup \hat{i}$
            \STATE $B = B + \lambda_{\hat{i}}$
            \STATE $n=n+1$
		\ENDIF
    \ENDWHILE
	\FORALL {$i \in \mathcal{R}$}
	    \STATE GET $\upsilon_i$	
	    \STATE UPDATE $\bar\theta_i(t)$	
	    \STATE $W_i = W_i +1$
	\ENDFOR
    \STATE UPDATE $\mathcal{L}(t)$
    \STATE $B=0$; $n=0$
\ENDFOR
\end{algorithmic}
\end{footnotesize}
\end{algorithm}

\subsubsection{ONETS: Online NETwork Slice broker}
While the \emph{eUCB} algorithm achieves outstanding performance, it requires solving an Integer Linear Programming (ILP) problem at every round and thus, it is not a feasible solution in practice. UCB applied to the Budget-Limited Multi-Armed Bandit problem has been proven to be NP-HARD in~\cite{knapsack_mab}. 
Therefore we designed a lower complexity version of \emph{eUCB}, namely \textbf{ONETS}, where exactly $K$ arms are selected every round while meeting the budget constraint. Clearly, this modification drastically reduces the computational time to $O(K)$ (see Alg.\ref{alg:ucb_k}) but at the performance cost of a sub-optimal solution (both effects are studied in Section~\ref{s:validation}). 

The pseudocode is listed in Alg.~\ref{alg:ucb_k}. The idea is to substitute Problem~\texttt{D-ONLINE-SLICING} with a practical procedure: Indexes $\hat{i}$ with active lock-up sessions are included in the next round (Line $9-10$), as the budget constraint has been already fulfilled in the previous round. If those indexes are not enough (less than $K$), we search the index among the remaining ones such that the empirical distribution mean is maximized while fitting into the budget left (line $14$).

\subsubsection{Regret Upper bound for ONETS}
\label{s:ubEUCB}

We provide an upper bound analysis for the \emph{ONETS} scheme. Let us consider  set $\hat\Theta = \{\hat\theta_i\}$ of empirical distribution means per arm $i$ obtained within time window $T$, where $\hat\theta_i = \bar\theta_i + G_{\text{ALG}}$ $\left( \text{in case of \emph{eUCB}} \text{, G}_{\text{ALG}} = \sqrt{\frac{2\log T}{W_i(T)}}\right)$. We can calculate the expected number of times arm $i$ is pulled down based on \emph{ONETS} algorithm as follows.
\begin{equation}
\label{eq:prob}
\begin{array}{l}
\mathbb{E}[W_i(T)] \leq \\
\bigintsss\limits_0^\infty f(x,\hat\theta_i) \text{Pr}(x \geq \max\limits_{p\neq i} \upsilon_p) \text{d}x + \\
\bigintsss\limits_0^\infty f(x,\hat\theta_i)  \underset{\forall j \in \mathcal{I}}{\text{Pr}}(\text{single } \upsilon_j \geq x) \text{Pr}(x \geq \max\limits_{p\neq \{i,j\} } \upsilon_p) \text{d}x + \\
\bigintsss\limits_0^\infty f(x,\hat\theta_i) \underset{\forall j \in \mathcal{I}}{\text{Pr}}(\text{multiple } \upsilon_j \geq x) \text{Pr}(x \geq \max\limits_{p\neq i} \upsilon_p) \text{d}x \leq \\
\sum\limits_{|\mathcal{H}|=0}^{|\mathcal{K}|}\!\!\sum\limits_{\sigma=1}^{{|\mathcal{I}|}\choose{|\mathcal{H}|}}\!\!\bigintsss\limits_0^\infty\!\! f(x,\hat\theta_i) \!\!\left(\!\! \prod\limits_{j\in \mathcal{H}(\sigma)}\!\! \bigintsss\limits_x^\infty\!\! f(y,\hat\theta_j) \text{d}y\!\! \prod\limits_{\substack{k\in\mathcal{I},\\ \mathcal{I}\setminus \{i\}\\ \mathcal{I}\setminus\mathcal{H}(\sigma)}}\!\!\!\! \bigintsss\limits_0^x\!\! f(y,\hat\theta_k) \text{d}y\!\!\right)\!\!\text{d}x.
\end{array}
\end{equation}

Considering a negative exponential distribution\\
$f(x,\hat\theta_i) = \frac{e^{-\frac{x}{\hat\theta_i}}}{\hat\theta_i}$ and recalling that $\int\limits_0^x f(x,\hat\theta_i)\text{d}x = 1-e^{-\frac{x}{\hat\theta_i}}$\\
we can obtain the following upper bound for $\mathbb{E}[W_i(T)]$
\begin{equation}
\label{eq:upbound}
  \sum\limits_{|\mathcal{H}|=0}^{|\mathcal{K}|-1}\sum\limits_{\sigma=1}^{{|\mathcal{I}|}\choose{|\mathcal{H}|}} \sum\limits_{\substack{\phi\in \\ \powerset(\mathcal{I}\setminus\{\mathcal{H}(\sigma)\},\\\mathcal{I}\setminus \{i\})}}\!\!\!\!\!\!\! (-1)^{|\phi|} \frac{1}{\hat\theta_i \left(\frac{1}{\hat\theta_i} + \!\!\!\sum\limits_{j\in\mathcal{H}(\sigma)} \frac{1}{\hat\theta_j} + \sum\limits_{p\in\phi} \frac{1}{\hat\theta_p}\right)};
\end{equation}
where $\sigma$ is the permutation (index) of all elements included in $\mathcal{H}$, while $\powerset(\mathcal{I})$ is the power set of all elements included in $\mathcal{I}$. 
Note that, in this paper, the reward distribution function can be approximated to a negative exponential distribution as we assume an exponential distribution for the arrival time of slice requests. However, complex distributions can be used to derive advanced upper bounds. 
For the sake of brevity, we leave to the reader the derivation  from Eq.~\eqref{eq:prob} to Eq.~\eqref{eq:upbound}. 
Also note that, the first two summations (on the left of Eq.~\eqref{eq:upbound}) can be also expressed as $\powerset({K})$, however we prefer to explicitly keep them to provide the computational effort for solving that equation.
By substituting the empirical distribution mean $\hat\theta_i$ with the value provided by the algorithm, assuming $L_i << T$ and using Eq.~\eqref{eq:regret}, we can obtain the upper bound of the regret, i.e., $\mathbb{E}[R_T^{\pi}(\Theta)] = O(log(T))$ if we apply \emph{ONETS}. When $T \sim L_i$, $\mathbb{E}[R_T^{\pi}(\Theta)] = O(log(T)+L^*)$, where $L^* = \max\limits_{i\in\mathcal{I}} L_i$, as also confirmed in~\cite{lookups}.

\subsection{$\epsilon$-greedy algorithm}
A greedy solution is the simplest algorithm for approaching BLMAB problems, where $K$ is not needed. It implies that the balance between exploitation and exploration is driven by a random $\epsilon$ value.
A linear dependency of $\epsilon$ with the elapsed time $t$ helps the selection policy to explore ``more'' neighboring solutions during the first rounds (as inferred distribution means might not be accurate) while ``trusting'' more on known distributions along the evolution of the experiments. We set $\epsilon=\frac{b|\mathcal{I}|}{d^2t}$ where $d\in \{0,1\}$ and $b>0$ are arbitrary values, as shown in Section~\ref{s:validation}. 
The algorithm (listed in Alg.~\ref{alg:greedy}) will select the best arms maximizing the reward (line $9$) with probability $\epsilon$, whereas it will select arms randomly (line $11$) satisfying (in both cases) the limited budget constraint (lines $14-17$). Please note that $\epsilon$-greedy algorithm does not require an initial training phase. 
\begin{algorithm}[t!]
\caption{\small{\textbf{$\epsilon$-greedy:} Selection algorithm $\pi$ to select the next batch of arms based on $\epsilon$-exploration probability.}}
\label{alg:greedy}
\algsetup{indent=1em}
\begin{footnotesize}
\begin{algorithmic}
	\STATE {\bf Input: }$ \mathcal{I}, T, C, b, d$
	\STATE {\bf Initialization: } $W_i = 0, \bar\theta_i(0) =0 \in \bar\Theta; \forall i\in\mathcal{I}, \mathcal{L} \leftarrow \emptyset$
	\STATE {\bf Procedure}
	\end{algorithmic}
	\begin{algorithmic}[1]
\FORALL {$t \in T$}
    \STATE $\epsilon = \min\{1,\frac{b|\mathcal{I}|}{d^2t}\}$
	\WHILE{$(C-B\geq 0) or (\mathcal{I}\neq\emptyset)$}
    	\IF{$\mathcal{L}(t) \neq \emptyset$}
        	\STATE $\hat{i} \leftarrow \mathcal{L}$
        \ELSE
        	\STATE GET $z\in[0,1]$ (uniformly distributed)
            \IF{$z>\epsilon$}
            	\STATE $\hat{i}: \arg\max\limits_{\mathcal{I}\setminus\mathcal{L}} \bar\theta_i(t)$
            \ELSE
    			\STATE $\hat{i}: rand(\mathcal{I}\setminus\mathcal{L})$
            \ENDIF
        \ENDIF
        \IF{$B + \lambda_i \leq C$}
			\STATE $\mathcal{R} \leftarrow \mathcal{R} \cup \hat{i}$
            \STATE $B = B + \lambda_i$
		\ENDIF
    \ENDWHILE
	\FORALL {$i \in \mathcal{R}$}
	    \STATE GET $\upsilon_i$	
	    \STATE UPDATE $\bar\theta_i(t)$	
	    \STATE $W_i = W_i +1$
	\ENDFOR
    \STATE UPDATE $\mathcal{L}(t)$
    \STATE $B=0$
\ENDFOR
\end{algorithmic}
\end{footnotesize}
\end{algorithm}

\subsubsection*{Regret Upper Bound for $\epsilon$-greedy}

Greedy solutions are proved to have a sub-linear regret. In particular, as shown in~\cite{AuerMab2002} the upper bound regret for $T\rightarrow\infty$ is expressed as $b/(d^2T)+o(1/T)+O(1/T^{1+\epsilon})$. 
Since our lock-up period constraint might only affect the number of times sub-optimal arms are randomly selected, i.e., $L^* = \max\limits_{i\in\mathcal{I}} L_i$, such an upper bound works also for our BLMAB considering that the probability to select a sub-optimal arm is at most the following:
\begin{equation}
\begin{array}{l}
\underset{i\neq i^*}{P}\{a_{i,t}\!\!=\!\!1\}\!\!=\!\! \frac{b}{d^2t}\!+\!\!2\!\left(\frac{b}{d^2}\log\frac{(t-1)d^2e^{1/2}}{b|\mathcal{I}|}\right)\!\!\left(\frac{b|\mathcal{I}|}{(t-1)d^2e^{1/2}}\right)\!\!^{\frac{b}{(5d^2)}} + \\ \frac{4e}{d^2}\left(\frac{b|\mathcal{I}|}{(t-1)d^2e^{1/2}}\right)^{b/2},
\end{array}
\end{equation}
where $i^*$ represents the optimal arm.

\section{Numerical Results}
\label{s:validation}

In this section, we validate our findings through numerical simulations carried out using a commercial tool, MATLAB\textsuperscript{\textregistered}.
In particular, we deploy a budgeted lock-up multi-arm bandit (BLMAB) problem as discussed in Section~\ref{s:blmab}. Network slice requests are generated following exponential distributions, as explained in Section~\ref{s:design}, with given $\rho,\zeta$ parameters. Every round, a selection policy is invoked to select a batch of tenants $K\geq 1$ to be granted. When a tenant is granted, it can (randomly) choose the slice template $s\in\mathcal{S}$ for its own traffic. Network traffic utilization is obtained in terms of used PRBs, as a sequence of i.i.d. random variables upper-bounded to the available number of slice template resources. 
All our simulations are run and results averaged over $1000$ random seeds to cope with randomness effects, providing a $95\%$ confidence degree.
The default system parameters are listed in Table~\ref{tab:system_param}. 


\begin{table}[h!]
\caption{Simulation parameters}
\label{tab:system_param}
\scriptsize
\centering
\begin{tabular}{|c|c|}
\hline
\textbf{System Parameters} & \textbf{Values}\\
\hline
$|\mathcal{I}|$ & $10$ tenants\\
$|\mathcal{S}|$ & $10$ slice templates\\
Capacity ($C$) & $150$ RBs ($15$ Mhz)\cite{ts23203}\\
$\rho ;\zeta$ & $100;0.1$\\
$\alpha$ & $0.5$\\
Time horizon $(T)$ & $10000$ rounds\\
$\epsilon$-greedy ($b;d$) & $10;0.01$\\
K (ONETS) & $6$\\
\hline
\end{tabular}
\end{table}

\begin{figure} [t]
\centering
	\includegraphics[width=0.45\textwidth]{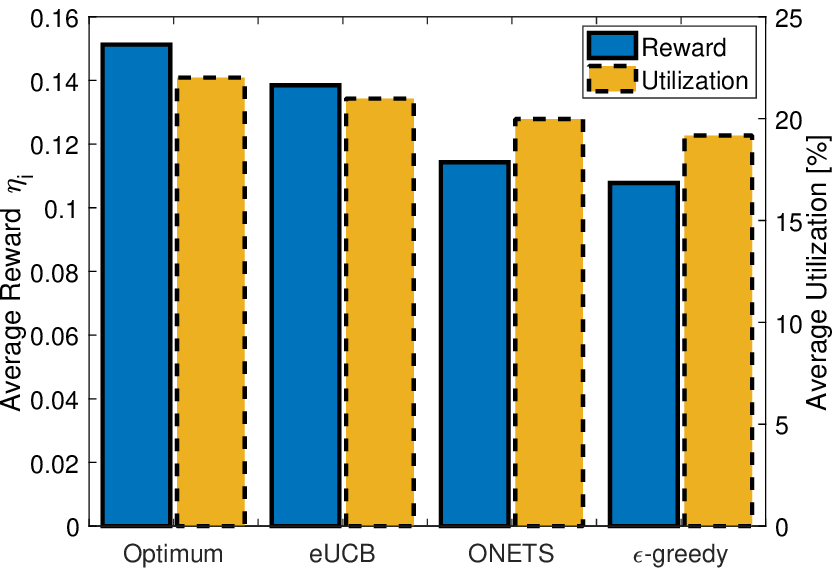}
	\caption{Performance evaluation of BLMAB Heuristics versus the \emph{Optimum} and \emph{eUCB} solutions.}
	\label{fig:optimal_perf}
\end{figure}

\subsection{BLMAB Solutions: Optimal vs Heuristics}

Fig.~\ref{fig:optimal_perf} shows a comparison of all solutions previously described for a network scenario with $5$ tenants, in terms of average reward $\bar{\eta_i}$ and average system utilization. As expected \emph{eUCB} is the closest solution to the \emph{Optimum} with \emph{ONETS} closely following it.

\begin{figure*}[t!]
    \centering
    \subfigure[Cumulative Rewards Distribution.]
    {
        \centering
        \label{fig:cdf_reward}
        \includegraphics[width=0.31\textwidth]{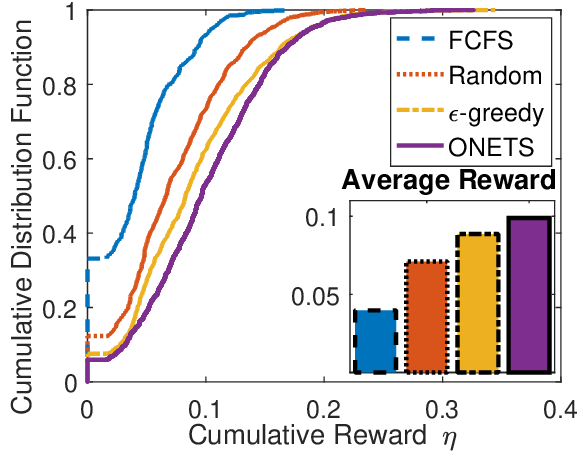}
    }
    \subfigure[Tenant Selection Ratio]
	{
        \centering
        \label{fig:sel_ratio}
        \includegraphics[width=0.31\textwidth]{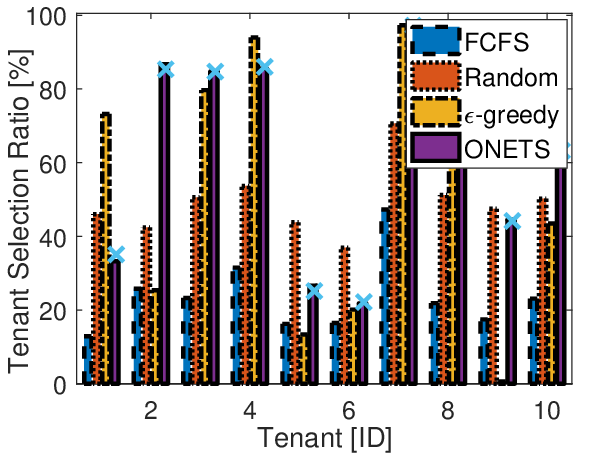}
    }
	\subfigure[System Utilization Distribution]
	{
        \label{fig:cdf_util}
        \centering
        \includegraphics[width=0.31\textwidth]{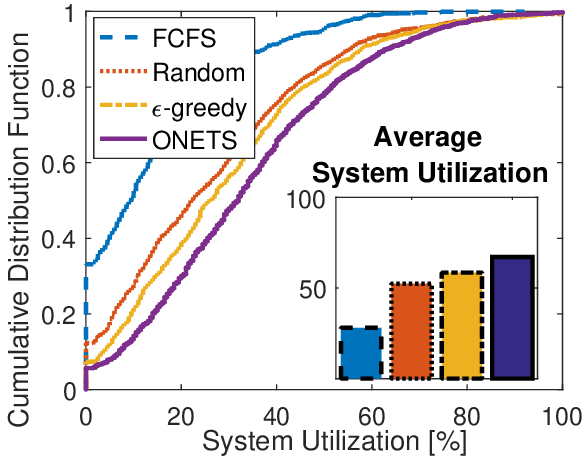}
    }
    \caption{Performance evaluation of different BLMAB selection policies for a 10 tenants scenario.}
    \label{fig:comparison}
\end{figure*}

\emph{Optimum} results have been obtained through the commercial tool IBM CPLEX OPL\textsuperscript{\textregistered} solver, used to solve Problem~\texttt{ONLINE-SLICING}, whereas \emph{eUCB} results are retrieved after solving Problem~\texttt{D-ONLINE-SLICING} for any single time $t$. \emph{ONETS} is run with $K$ empirically set to $K = 3$. 


Table~\ref{tab:computation} shows the computational counterpart to the average reward and utilization results. We show the measured computational time for running every solutions for a single instance (referred as Inst.) as well as for the whole simulation period of $1000$ rounds (referred as Sim.). As it can be observed, the performance gains previously observed from the \emph{Optimum} and \emph{eUCB} solutions come at a poor scalability with the number of tenants deeming them as unfeasible in practice. In the rest of the experiments we will not consider them anymore for feasibility reasons. 

\begin{table}[h!]
\caption{Computational Load}
\label{tab:computation}
\small
\centering
\begin{tabular}{| c || c | c | c | c | c | c |}
\hline
\bf{Solutions} & \multicolumn{2}{c|}{{\bf 5 Tenants}} & \multicolumn{2}{c|}{{\bf 10 Tenants}} & \multicolumn{2}{c|}{{\bf 15 Tenants}}\\
\hline
& Sim. & Inst. & Sim. & Inst. & Sim. & Inst.\\
\hline
Optimum & \multicolumn{2}{c|}{$24109$ s} & - & - & - & -\\
eUCB & $4708$ s & $131$ s & $11512$ s & $543$ s & - & - \\
$\epsilon$-greedy & $219$ s & $0.3$ s & $398$ s & $0.5$ s & $502$ s & $0.6$ s\\
\bf{ONETS} & $\mathbf{322}$ \bf{s} & $\mathbf{1.2}$ \bf{s} & $\mathbf{501}$ \bf{s} & $\mathbf{1.9}$ \bf{s} & $\mathbf{847}$ \bf{s} & $\mathbf{2.6}$ \bf{s}\\
\hline
\end{tabular}
\end{table}

\subsection{BLMAB Heuristics Benchmarking}
Given the lack of existing solutions in this particular context, we consider for benchmarking purposes two baseline approaches. First, we consider a trivial selection policy, First Come First Served (\emph{FCFS}), to accept all incoming network slice requests as far as there are enough resources in our network. Second, we consider a random selection policy process (\emph{Random}) that chooses tenants (or subset of tenants) based on a uniform distributed random variable, while satisfying budget constraints.


\subsubsection*{Arms Selection Policies}
In Fig.~\ref{fig:comparison} we depict three key performance figures for a $10$ tenants scenario.  Fig.~\ref{fig:cdf_reward} shows the differences obtained by the different approaches in terms of reward. Fig.~\ref{fig:sel_ratio} shows the number of times each tenant $i$ is selected by the different approaches. Note that inter-arrival times $\Delta t$ of slice requests per tenant are exponentially distributed with rate $\phi_i$ (see Section~\ref{s:design}). On the same figure, we also plot with cross signs the expected number of times each tenant is selected based on Eq.~\eqref{eq:prob} as described in Section~\ref{s:online}. This result supports our model, as our analysis accurately predicts the number of times each tenant is selected by \emph{ONETS}.
Fig.~\ref{fig:cdf_util} shows the system utilization percentage achieved with the different approaches.

Based on these results \emph{ONETS} fulfills its design objectives outperforming at different levels the different alternative approaches.

\subsubsection*{Number of Tenants}
\begin{figure} [!t]
\centering
	\includegraphics[width=0.5\textwidth]{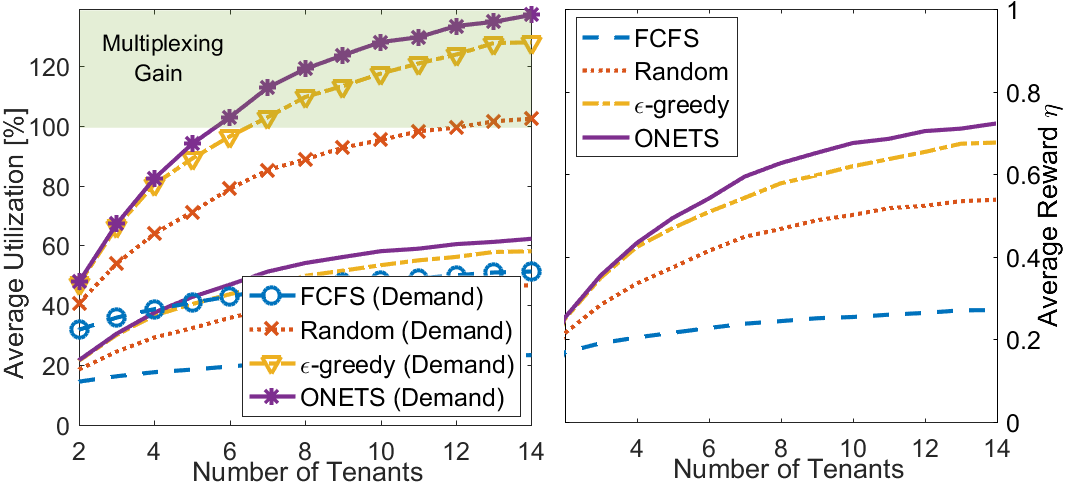}
	\caption{Performance evaluation of different BLMAB selection policies when increasing the number of tenants.}
	\label{fig:increas_perf}
\end{figure}

In Fig.~\ref{fig:increas_perf}, we study two performance metrics as the number of tenants is increased. On the left picture, we show the average utilization for all mechanisms compared with the total slice resource demand. While increasing the number of tenants intuitively leads to a higher average utilization, it also benefits the potential multiplexing gain as shown in the green area. On the right side, we show the average reward while increasing the number of tenants. \emph{ONETS} outperforms the other solutions showing consistent gains in terms of reward, average utilization and multiplexing gain. 

These results suggest that operators would benefit of ``opening'' their networks to external tenants through network slicing given the potential gains in increasing the overall system utilization and corresponding profit.

\begin{figure*}[t!]
    \centering
    \subfigure[$\alpha=0.1$]
    {
		\label{fig:alpha_1}        
        \centering
        \includegraphics[width=0.31\textwidth]{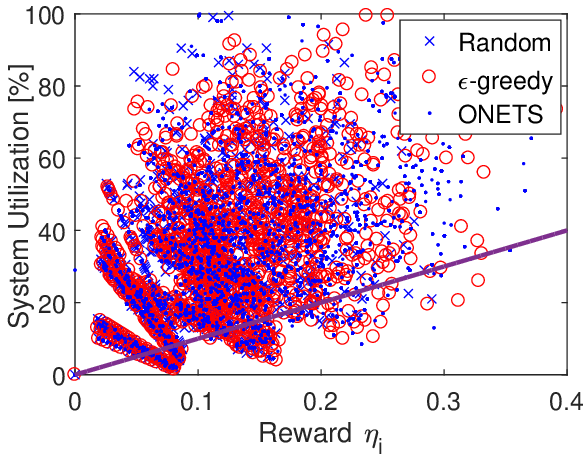}
    }
    \subfigure[$\alpha=0.5$]
	{
		\label{fig:alpha_5}        
        \centering
        \includegraphics[width=0.31\textwidth]{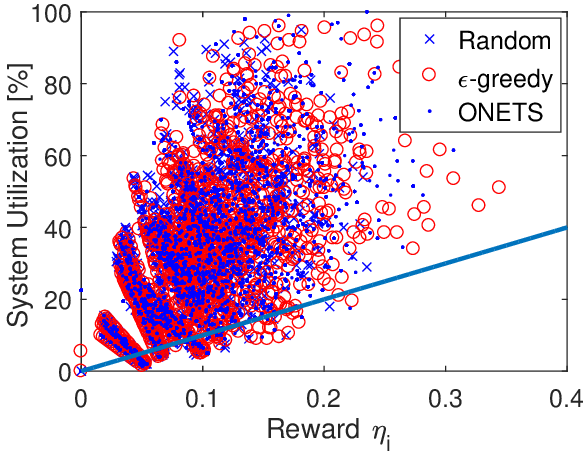}
    }
	\subfigure[$\alpha=0.9$]
	{
		\label{fig:alpha_9}        
        \centering
        \includegraphics[width=0.31\textwidth]{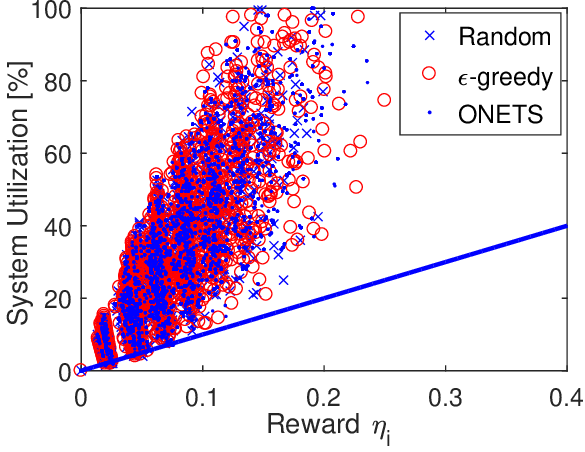}
    }
    \caption{Scatterplot of Reward versus System Utilization for different $\alpha$ configurations}
    \label{fig:scatterplot}
\end{figure*}

\subsubsection*{Network Slicing Multiplexing Control ($\alpha$)}

In this section, we study our reward solution expressed in Eq.~\eqref{eq:reward} in order to illustrate the impact of the configurable parameter $\alpha$. In particular, we show its behaviour in relation to system utilization. Fig.~\ref{fig:scatterplot} shows a scatter plot of reward obtained per round compared with the system utilization achieved at that particular time, when different solutions are applied. The overall behavior suggests a strong dependency of system utilization from the game reward $\eta_i$, which validates our reward design: the larger the reward, the higher the multiplexing gains. 

However, a proper tuning of the weight $\alpha$ might strongly influence and lead the system toward near-optimal steady states. When $\alpha$ equally distributes the weight (Fig.~\ref{fig:alpha_5}), the system efficiency is equally distributed around the linear dependency line (dashed line). When $\alpha=0.9$ in Fig.~\ref{fig:alpha_9}, the strong dependency results in a strong perturbation of results: a small variation of reward may cause a significant increase of utilization resulting in an unstable behavior when performed in an online fashion. All results show that \emph{ONETS} is the best scheme efficiently translating higher rewards to higher multiplexing gains. 

\subsubsection*{SLA Protection vs Multiplexing Gains}
\emph{ONETS} relies on the BLMAB framework ability to predict traffic behaviour based on past observations. However, outliers (i.e., traffic bursts) might lead to performance degradation and, in the worst case, to SLA violations. 

\begin{figure} [!t]
\centering
	\includegraphics[width=0.5\textwidth]{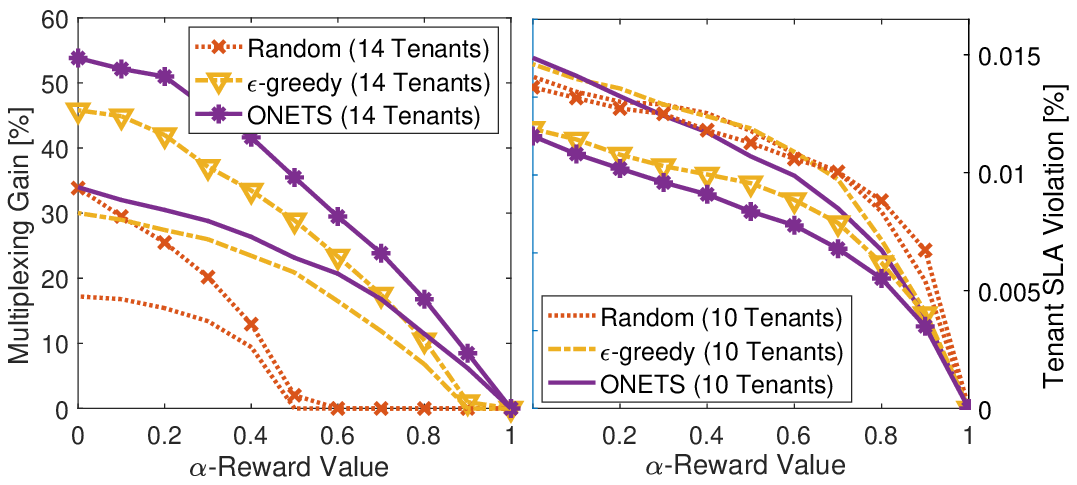}
	\caption{Performance evaluation of different BLMAB selection policies by varying the $\alpha$ configuration.}
	\label{fig:violations}
\end{figure}

In Fig.~\ref{fig:violations}, we show the network slicing multiplexing gains versus the average SLA violation per tenant computed as the percentage of the number of times slice resources were not fully provided to tenants divided by the total number of slices granted. As it can be observed, our solution achieves high multiplexing gains ($>30\%$) at limited SLA violation risk ($<0.015\%$). 

This trade-off is optimally driven by a proper tuning of the reward $\eta_{i,t}$, expressed in Eq.~\eqref{eq:reward}\,\footnote{We assume the same reward model for the whole system. Advanced reward models differentiating customers classes are out of scope of this paper and might be considered in future extensions.}. Specifically, different $\alpha$ values might strongly influence the overall system behavior, as shown in Fig.~\ref{fig:violations}. In case of $\alpha = 1$, only network slice request information is considered without any past information on the real slice utilization. This ensures no SLA violation but also no multiplexing gains. As $\alpha$ increasingly approaches to $0$, past traffic information is considered in our model, allowing for resource over-provisioning and thus, multiplexing gains. However, this comes at the cost of an increasing SLA violation risk.

\section{Proof-of-Concept}
\label{s:implement}
%
\begin{figure*}[!t]
\label{fig:poc}
\centering
	\subfigure[Testbed blocks overview]
    {
        \label{fig:poc_blocks}
        \includegraphics[clip, trim = 2.5cm 13cm 17.5cm 7cm, width=0.38\textwidth]{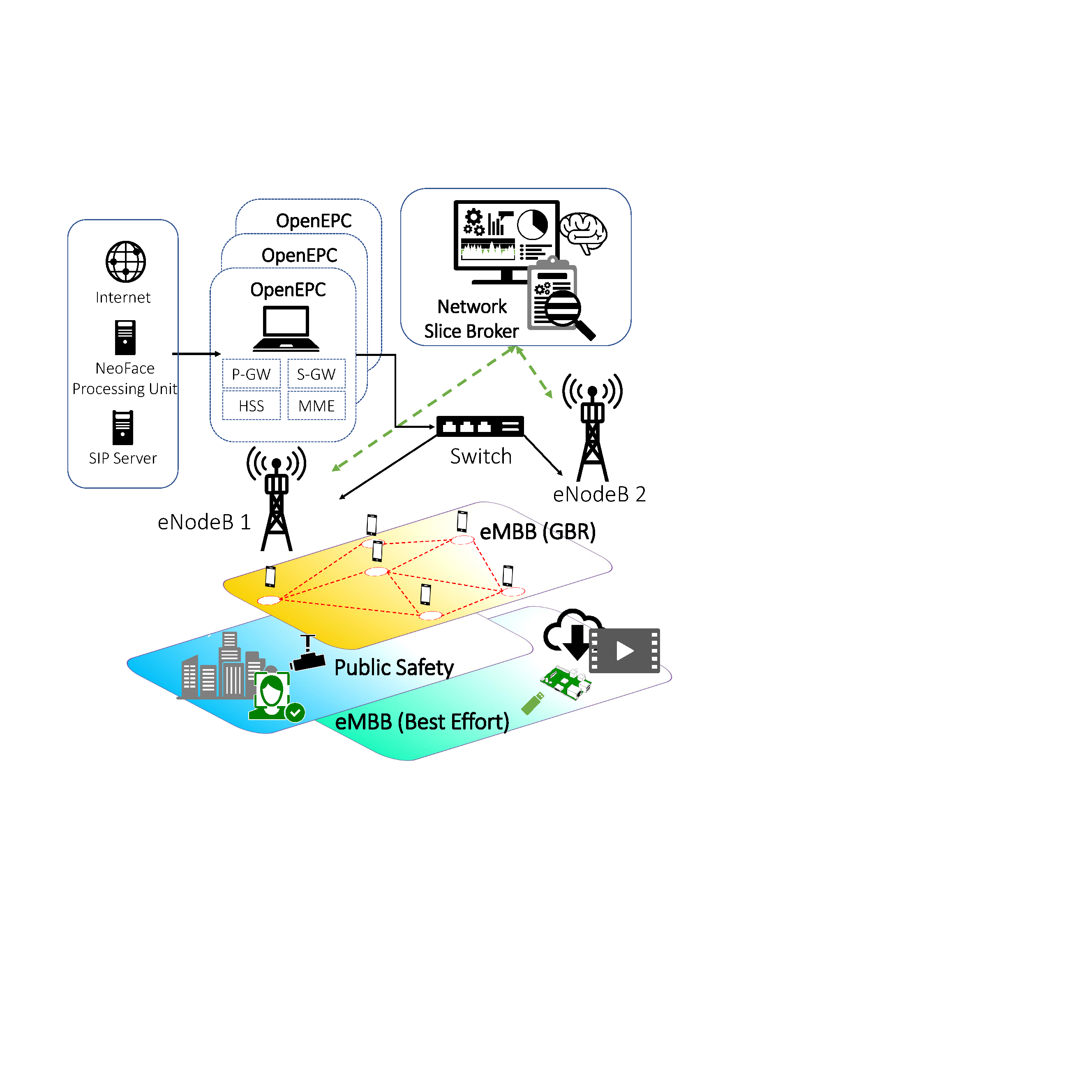}
    }
    \subfigure[Real deployment]
    {
        \label{fig:poc_photo}
        \includegraphics[clip, trim= 0cm 0cm 0cm 0cm ,width=0.48\textwidth]{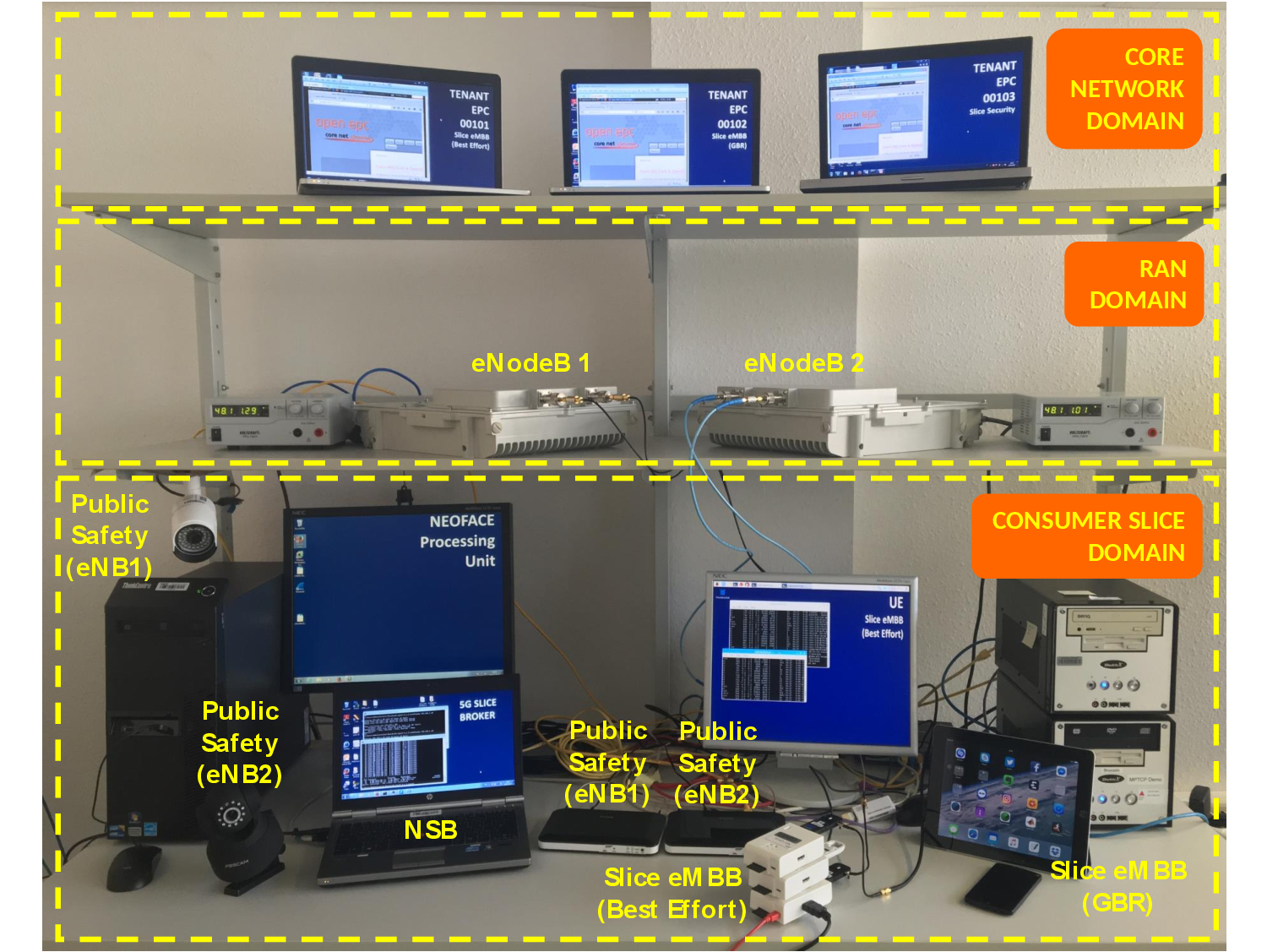}
    }
\caption{Network Slicing Proof-of-Concept with 3 Slices: eMMB (BE), Public Safety and eMBB (GBR).}
\end{figure*}
\subsection{PoC Setup}
In this section we describe the proof-of-concept implementation of our proposed \emph{ONETS} solution. We built on available commercial hardware to setup a testbed comprising: ($i$) three virtualized Evolved-Packet-Core (EPC) (one per slice), ($ii$) 2 LTE eNBs connected to the EPCs, ($iii$) multiple LTE devices generating traffic with different service requirements, such as mobile phones, surveillance cameras and USB dongles, ($iv$) our \emph{ONETS} solution implemented as a stand-alone software connected to the Local Maintenance Terminal (LMT) of the RAN environment. All equipment hardware specifications are listed in Table~\ref{tab:hardware}.

\begin{table}[b!]
\caption{PoC Hardware Specifications}
\label{tab:hardware}
\centering
\begin{tabular}{|c|c|c|}
\hline
\textbf{Equipment} & \textbf{Description} & \textbf{Ref.}\\
\hline
Virtualized EPC & OpenEPC Rel. $6$ & \cite{openepc}\\
Face Recognition SW & NeoFace Facial Recognition & \cite{neoface}\\
Surveillance Camera & HD, Motion Detection & \cite{surv_camera}\\
UE & LTE USB Stick & \cite{huawei-dongle}\\
Smart phones & iPhone (SE) and LG Nexus (5) & -\\
Router LTE & LTE with SIM-card slot & \cite{lte-router}\\
Smart devices & Raspberry Pi 2 &\cite{raspberry} \\
LTE Small cell & $15$ MHz channel, LTE Band $3$ & \cite{nec-smallcell}\\
& DL $1775$ MHz, UL $1870$ MHz (FDD)& \\ 
\hline
\end{tabular}
\end{table}

The OpenEPC software~\cite{openepc} contains all the functional elements of the 3GPP EPC up to Release 12. We deploy three separated virtual machines running on the same host machine. It automatically builds main LTE core network elements, such as HSS, AAA, S-GW, MME and P-GW. All the interfaces among them are virtualized through a hypervisor, VMWare Workstation. The host machine is provided with two external Gigabit ethernet interfaces: the former is used for an internet gateway connection, the latter is used for establishing the S$1$ interface with the RAN nodes.
Regarding the commercial eNBs~\cite{nec-smallcell}, they use 15MHz bandwidth, i.e., $150$ PRBs per $1$ms subframe (cfr. \cite{ts23203}). For practical reasons we carry out conducted tests: We abate wireless channel uncertainty providing the devices with CRC/SMA cables directly connected to the radio interface of the eNBs.

To support the network slicing concept, we build on RAN equipment supporting RAN Sharing: This enables us to use the same RAN infrastructure for different Public Land Mobile Networks (PLMNs). We apply the Multiple Operator Radio Access Network (MORAN) approach to have dedicated network core domains sharing the same RAN facilities. Each UE connects through the same set of eNBs indicating the PLMN-id, i.e., the slice id, for being served. In our experiments, a fixed number of mobile cores is already instantiated (corresponding to the number of tenants in our system). When a network slice request is accepted, eNBs are dynamically configured to activate an additional PLMN-id and to route traffic associated with users under this PLMN-id to its dedicated MME (and virtualized EPC network).

%
%

We assume three different tenants registered for issuing network slice requests with different traffic characteristics: ($i$) enhanced Mobile BroadBand (eMBB BE) slice generating FTP file transfers best-effort traffic, ($ii$) eMBB GBR generating multiple Voice over LTE (VoLTE) traffic streams emulating an audio conference system, ($iii$) a Public Safety slice for video surveillance.

Raspberry Pis~\cite{raspberry} with LTE USB dongles~\cite{huawei-dongle} are used to generate eMBB BE traffic. Tenants might ask for a new network slice only if an own network slice is not already running. Our network slice broker dynamically receives network slice requests and at run-time decides whether to accept the network slice and configure the eNBs accordingly. eNBs provide an LMT interface to properly tune the number of physical resource blocks (PRBs) assigned per PLMN. Once the slice is accepted and correctly instantiated, the network slice broker monitors the slices traffic to retrieve statistical information for future network slicing decisions. An overview of the system is depicted in Fig.~\ref{fig:poc_blocks}. In the eMBB GBR slice case, commercial cellular phones and tablets are used for generating voice traffic.

As one of the slices in the testbed is tailored to public safety purposes, we deploy surveillance cameras in our testbed. We connect the IP cameras to LTE routers~\cite{lte-router}, which are, in turn, connected to our eNBs. The cameras~\cite{surv_camera} are provided with motion detection features. This introduces a bursty traffic source for our experiments. In addition, an advanced face recognition software~\cite{neoface} is fed with video streaming traffic to detect face recognition matches to faces stored in a database. 

\begin{figure} [!t]
	\centering
	\includegraphics[clip, trim = 2cm 27cm 22cm 0cm, width=0.5\textwidth]{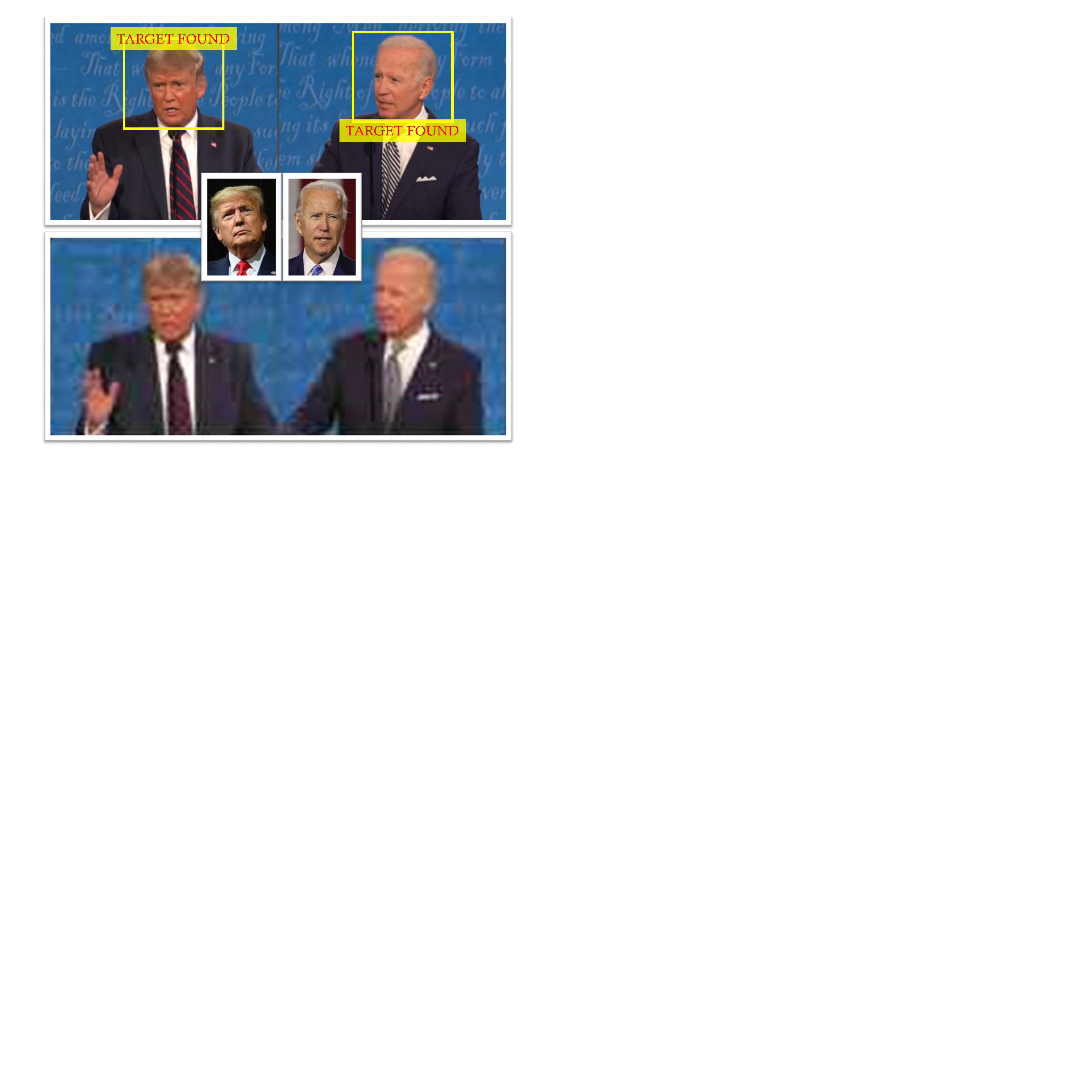}
	\caption{Sample of video surveillance face detection: successful (upper side) and failed (lower side).}
	\label{fig:Neoface}
\end{figure}

The recognition server is attached to the P-GW through the SGi interface. A target list is already loaded in the recognition server, 
When the face recognition software matches a known target, a yellow square appears around the face detected (see Fig.~\ref{fig:Neoface}). However, a detection threshold parameter may affect the detection process. Indeed, if the quality of the video stream is below a pre-determined threshold, the detection process may fail and the target might not be correctly recognized. The images quality is dynamically adjusted based on the LTE channel condition. 

The IP camera stores video streams while a VLC server dynamically encodes the live-video based on the channel quality feedback from the recognition server.
In Fig.~\ref{fig:poc_photo} the different proof-of-concept components are depicted, where we highlight the service domain (Consumer slice domain), the RAN infrastructure (RAN domain) and the core network domain where the three EPCs are deployed.

\subsection{PoC Performance evaluation}
\label{s:results}

We evaluate our \emph{ONETS} online network slicing solution in the proof-of-concept setup previously described. Our goal is to analyze the feasibility of our approach and get insights on the potentially achievable gains, i.e., exploit the multiplexing gain between the amount of resources assigned to a particular network slice and the actual slice resources utilization in time. 

\begin{figure} [!t]
\label{fig:testbed1}
\centering
    \subfigure[eMBB (BE) Slice]
    {
        \label{fig:testbed1b}
        \centering
        \includegraphics[width=0.5\textwidth]{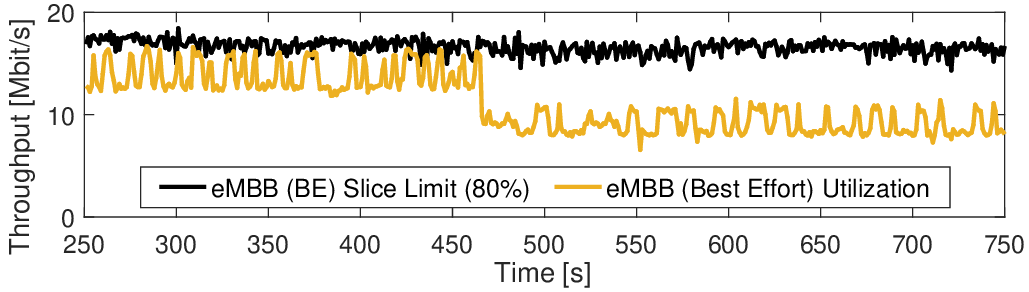}
    }
    \subfigure[Public Safety Slice]
    {
        \label{fig:testbed1c}
        \centering
        \includegraphics[width=0.5\textwidth]{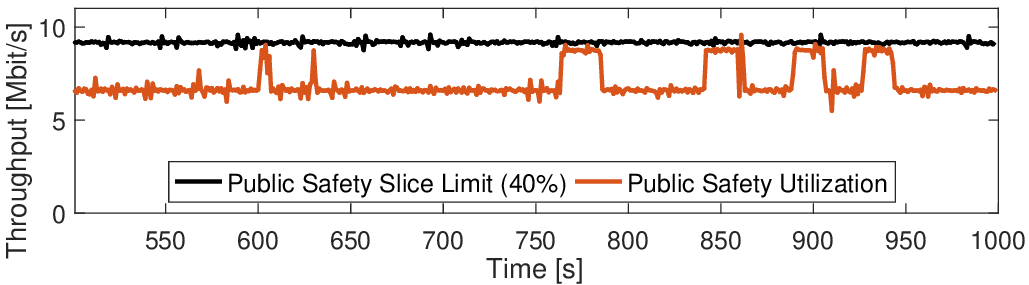}
    }
	\subfigure[eMBB (GBR) Slice.]
    {
        \label{fig:testbed1a}
        \centering
        \includegraphics[width=0.5\textwidth]{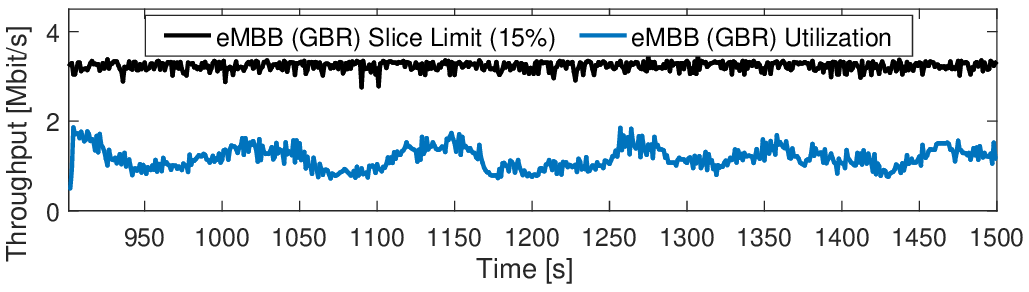}
    }
	\caption{Network Slices Traffic: Granted vs Measured}
	
\end{figure}

We consider three slices that are sequentially introduced in our system and dimensioned for peak demand.
First, we introduce a baseline eMBB (Best-Effort) slice requesting $80\%$ of the system capacity for FTP transfers. Fig.~\ref{fig:testbed1b} depicts the traffic pattern in time of this slice and the difference to the granted slice limit.
Second, we introduce a Public Safety slice for video surveillance, see Fig.~\ref{fig:testbed1c}, where two surveillance cameras upload video streams for face recognition within a $1500$ seconds time window and illustrate the difference to the granted traffic slice limit ($40\%$ of the system capacity). 
Finally, we introduce an eMBB (GBR) slice for audio conferences. We considered $30$ devices generating voice calls.
Each device is provided with a custom SIM-card, configured in our core domain to belong to a single PLMN-id, i.e., a single network slice. eNBs are configured to allocate an eMBB (GBR) network slice with $15\%$ of system capacity as demand. Fig.~\ref{fig:testbed1a} shows the measured utilization versus the granted slice limit. 

\begin{figure}[t!]
	\label{fig:testbed}
    \centering
    \subfigure[Network slices aggregated offered load]
    {
        \label{fig:testbed3a}
        \centering
        \includegraphics[width=0.5\textwidth]{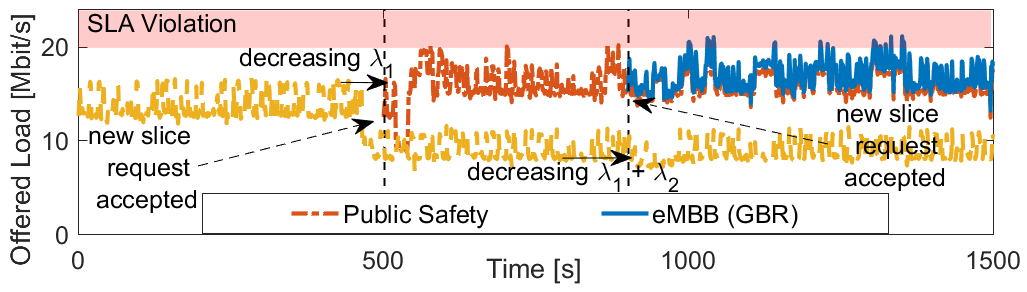}
    }
    \subfigure[System utilization]
	{
        \label{fig:testbed3b}
        \centering
        \includegraphics[width=0.5\textwidth]{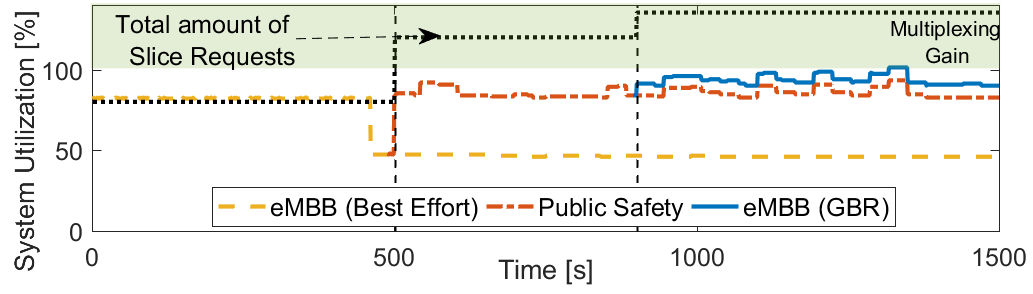}
    }
    \subfigure[System reward]
	{
        \label{fig:testbed3c}
        \centering
        \includegraphics[width=0.5\textwidth]{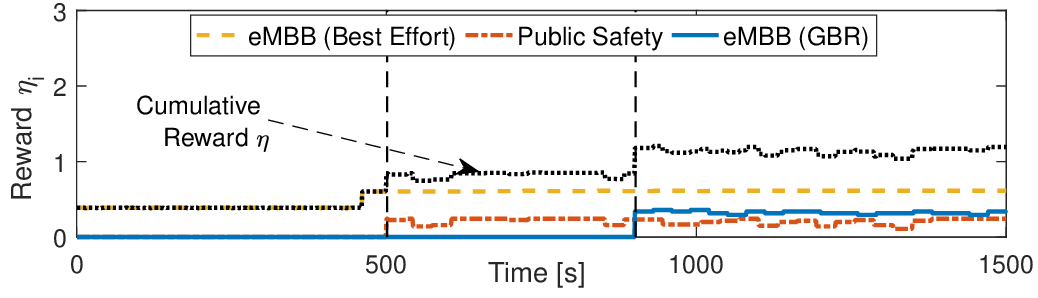}
    }
    \caption{\emph{ONETS} Online Network Slice Broker in action sequentially granting network slices.}
\end{figure}

In Fig.~\ref{fig:testbed3a}, we show the dynamic system behavior when \emph{ONETS} is applied. The eMBB (Best-Effort) network slice is dimensioned to use $80\%$ of the system capacity. Initially, the slice is fully using its resource allocation. During this period, if new network slice requests arrive, the system might reject them if they are above the leftover capacity.

After $480$ seconds, we reduce the eMBB (BE) slice offered load by reducing the FTP file transfers. Our \emph{ONETS} solution automatically detects a change in the system utilization ($\lambda_1$) and triggers the selection policy to consider admitting new network slice requests. When the Public Safety network slice request arrives, the system checks its feasibility and allocates it resulting in a higher reward ($\eta_1 + \eta_2$), as shown in Fig.~\ref{fig:testbed3c}. 

After $900$ seconds, the traffic associated to the eMBB (BE) slice is decreased again. The system capacity variation is detected ($\lambda_1,\lambda_2$) and a new network slice admitted into our system: eMBB (GBR). As its traffic is scheduled, a higher system reward is achieved ($\eta_1 + \eta_2 + \eta_3$). The multiplexing gain is shown in Fig.~\ref{fig:testbed3b}, where the system utilization is compared with the aggregated granted network slice resources (the green area indicates a utilization above 100\%). 

As it can be observed, the network slicing multiplexing gains achieved with \emph{ONETS} allow for increasing the number of slices that can be accepted in the system. In this illustrative example network slice requests can be admitted up to $\approx$ 120\% of the system capacity, thus \emph{virtually} increasing the \emph{effective} capacity of the system and the achievable profit, accordingly. The cost of this gain is shown in 
Fig.~\ref{fig:testbed3a} where after admitting in the system the second and third slice request, there are peaks of offered load that hit the maximum available capacity and thus, the SLA protection level could be threatened. 

\section{Conclusions}
\label{s:concl}

One of the key novel concepts of 5G networks is Network Slicing, driven by use cases which are very diverse and sometimes with extreme requirements, e.g. automated driving, tactile internet, mission-critical. In this paper we proposed and analyzed \emph{ONETS}: an Online NETwork Slice broker solution that builds on the budgeted lock-up multi-armed bandit theory to design a low-complexity solution that maximizes network slicing multiplexing gains, achieving the accommodation of network slice requests in the system with an aggregated level of demands above the available capacity. 

Our results show that \emph{ONETS} ($i$) is feasible in practice as it has been successfully implemented and tested on top of a commercial LTE system, ($ii$) the achievable multiplexing gains are significant and increase according to the number of slices in the system, ($iii$) \emph{ONETS} clearly outperformed na{\"i}ve or greedy solutions (FCFS, Random, $\epsilon$-greedy) for the considered scenarios, ($iv$) its computational complexity is in the same order of magnitude of a simple greedy solution ($\epsilon$-greedy), ($v$) \emph{ONETS} aggressivity for achieving network slicing multiplexing gains is a configurable parameter ($\alpha$) that can be freely tuned by operators according to proprietary policies and desired SLA protection levels.






\bibliographystyle{IEEEtran}
{
\bibliography{bibliography}
}

\end{document}